\documentclass[reprint,aps,amsmath,amssymb,amsfonts,showkeys]{revtex4-2}
\usepackage[dvipdfmx]{graphicx}
\usepackage{dcolumn}
\usepackage{bm}
\usepackage{txfonts}
\usepackage{multirow}
\usepackage{color}

\begin{document}

\title{Carrier filtering effect for enhanced thermopower in a body-centered tetragonal ruthenate}

\author{Ryota~Otsuki}
\email{6223510@ed.tus.ac.jp}
\author{Yoshiki~J.~Sato}
\email{yoshiki\_sato@rs.tus.ac.jp}
\author{Ryuji~Okazaki}
\email{okazaki@rs.tus.ac.jp}
\author{Tomoya~Komine}
\author{Ryosuke~Kurihara}
\author{Hiroshi~Yaguchi}

\affiliation{Department of Physics and Astronomy, Tokyo University of Science, Noda 278-8510, Japan}

\begin{abstract}

Charged carriers in solids diffuse from hot to cold sides under temperature gradient to
induce the thermoelectric voltage.
Carrier filtering effect, which only passes either electrons or holes for the conduction process,
is an efficient method to enhance such voltage,
although it is challenging to experimentally realize it 
especially in conventional metals with
weak energy dependence of the density of states near the Fermi level.
Here we
measure the in-plane and out-of-plane thermopower of the layered perovskite Sr$_2$RuO$_4$ single crystals
above room temperature.
We find that the out-of-plane thermopower is largely enhanced with increasing temperature,
while the in-plane one seems to remain a temperature-independent constant value expected from the 
Heikes formula.
The observed large out-of-plane thermopower may originate from the 
recently proposed intriguing hole filtering effect in the body-centered tetragonal system,
in which the carrier hopping through the centered atom is essential.
Thus, the present carrier filtering effect may be a universal property
to be applicable in various materials belonging to such crystal system.

\end{abstract}

\maketitle

\section{introduction}

Thermoelectricity is a solid-state property to convert the 
heat current into charge one, or vice versa, 
attracting great attention
as an environmentally-friendly energy conversion technology \cite{He2017,Shi2020,Yan2022}.
In the fundamental point of view, 
the thermopower $S$,
which is the proportional coefficient between the electric field $\bm E$
and the temperature gradient $\nabla T$ as 
$\bm E = S \nabla T$,
serves an intriguing measure of how large the electron-hole asymmetry is in materials \cite{bbook};
under the temperature gradient,
the carrier diffusion from high- to low-temperature sides
induces the thermoelectric voltage, 
the magnitude of which is however significantly cancelled owing to 
the opposite polarities of the thermally excited electrons and holes.
To enhance the thermopower,
it is thus important to introduce a sort of asymmetry for the charged carriers:
for example, 
an energy barrier picture to pass only the unipolar carriers 
has been suggested in the superlattice structures based on the thermionic emission process \cite{Vashaee2004,Vineis2010}.
In the reciprocal space, 
a peculiar pudding-mold band shape to explain large thermopower in the cobalt oxides \cite{Terasaki1997,Kuroki2007,Usui2013}
is a kind of carrier filtering 
because it utilizes the large asymmetry in the carrier velocities 
to increase the thermopower and the electrical conductivity simultaneously.
Low-dimensional materials with highly asymmetric energy dependence of the density of states (DOS)
are also promising to enhance the thermoelectric efficiency \cite{Hicks1993} and has been experimentally demonstrated \cite{Ohta2007,Shimizu2016,Kouda2022}.
In a real-space picture, a spin blockade hopping in several cobalt oxides also 
clarifies the asymmetry in the thermoelectric transport \cite{Maignan2004,Taskin2005,Taskin2005a,Taskin2006,Chang2009,Negi2019,Okazaki2022}.
Moreover, the electron-hole asymmetry in the scattering rate 
is also crucial for the thermopower \cite{Georges2021,Gourgout2022}.

\begin{figure}[b]
\begin{center}
\includegraphics[width=7.5cm]{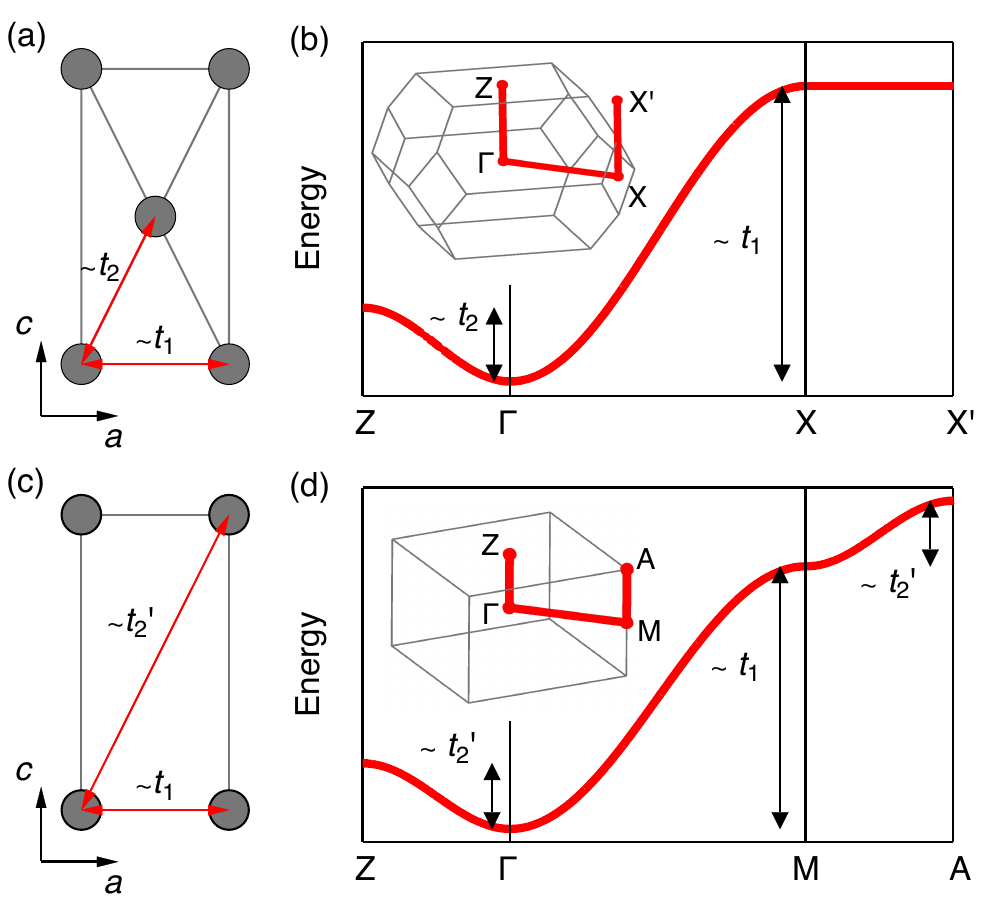}
\caption{
Concept of the carrier filtering in body-centered tetragonal lattice.
(a) Body-centered tetragonal lattice. 
(b) Tight-binding band dispersion for (a).
The inset depicts the Brillouin zone and the notations in Ref. \onlinecite{Haverkort2008} are used.
The out-of-plane velocity appears along the $\Gamma$-Z line
while it vanishes along the X-X' line.
The point X' denotes the X point in the adjacent Brillouin zone.
(c) Primitive tetragonal lattice. 
(d) Tight-binding band dispersion for (c). 
The out-of-plane velocity emerges in both the $\Gamma$-Z and the X-X' lines.
}
\end{center}
\end{figure}

For such carrier filtering effects, 
Mravlje and Georges have 
recently suggested unique filtering mechanism
based on the band structure in the body-centered tetragonal (bct) crystal system [Fig. 1(a)]
such as the layered ruthenate \cite{Mravlje2016}.
In a simple tight-binding picture,
the energy dispersion of the bct lattice in such layered materials is given as 
$\varepsilon(\bm k) = \varepsilon_{1}(\bm k) + \varepsilon_{2}(\bm k)$,
where 
$\varepsilon_{1}(\bm k) = t_1(\cos k_x+\cos k_y)$ ($t_1<0$) is 
the large in-plane hopping term 
and 
$\varepsilon_{2}(\bm k) =t_2\cos (k_x/2)\cos(k_y/2)\cos(k_z/2)$ ($t_2<0$, $|t_2| < |t_1|$)
represents the out-of-plane hopping one [Fig. 1(b)].
In this case, 
owing to the in-plane transfer integral $t_1$,
the energy along the $\Gamma$-Z line $(0,0,k_z)$ and 
the X-X' line $(\pm \pi,\pm \pi,k_z)$ yield
large negative and positive values, respectively.
The out-of-plane velocity 
along the $\Gamma$-Z line then becomes finite 
because of the dispersion $\varepsilon(0,0,k_z) = 2t_1 + t_2\cos(k_z/2)$,
while it becomes zero 
along the X-X' line
since 
$\varepsilon(\pm \pi,\pm \pi,k_z) = -2t_1$.
Therefore, 
the difference in the out-of-plane velocity 
along the $\Gamma$-Z and the X-X' lines
becomes  
significant.
This leads to the prominent electron-hole 
asymmetry when 
the chemical potential $\mu$ locates near the middle of the band, 
and the carrier filtering mechanism operates at high temperatures $T\sim|t_1|/k_{\rm B}$,
where the thermal energy $k_{\rm B}T$ is a substantial fraction of the bandwidth $|t_1|$.
Subsequently large thermopower is expected 
at such high-temperature range.
Note that
the situation is distinct from that of the primitive tetragonal structure with the 
out-of-plane hopping term of $\varepsilon_{2'}(\bm k) =t_2'\cos (k_x)\cos(k_y)\cos(k_z)$ [Figs. 1(c) and 1(d)].

The aim of this paper is to experimentally examine such carrier filtering effect 
in the proposed material Sr$_2$RuO$_4$.
The layered perovskite oxide Sr$_2$RuO$_4$ with the bct lattice (space group $I4/mmm$)
has been extensively studied as 
a model material to examine the unconventional superconductivity in two dimensions \cite{Maeno1994,Rice1995,Mackenzie2003,Kallin2012,Mackenzie2017,Kivelson2020},
while the pairing mechanism of the 
superconducting state is still a remaining issue \cite{Pustogow2019,Ishida2020,Chronister2021,Benhabib2021,Ghosh2021,Grinenko2021}.
Since the electronic structure and the quasi-two-dimensional (q-2D) Fermi surfaces
of this material have been well studied
both theoretically \cite{Oguchi1995,Singh1995,Hase1996,Haverkort2008,Gingras2019}
and experimentally \cite{Mackenzie1996,Yokoya1996,Damascelli2000,Shen2007,Tamai2019},
this material is indeed a minimal model for the examination of such filtering effect,
whereas the thermopower measurements on the single-crystalline samples of Sr$_2$RuO$_4$
have been limited in the low-temperature range \cite{Yoshino1996,Xu2008,Yamanaka2022}.

In the present study, 
we measure the in-plane and out-of-plane thermopower of Sr$_2$RuO$_4$ single crystals
at high temperatures.
The in-plane thermopower is well described within the Heikes formula that has been widely used to explain the high-temperature thermopower of correlated oxides. 
On the other hand, 
the out-of-plane thermopower increases with increasing temperature and exceeds the expected value of Heikes formula. 
These results indicate that the suggested carrier filtering effect is realized in the present bct system.   
Although the out-of-plane resistivity is relatively large in Sr$_2$RuO$_4$ \cite{Tyler1998},
this filtering is very universal for the bct lattice and acts at high temperatures,
offering a unique class of efficient thermoelectric materials.

\section{Methods}

Single crystals of Sr$_{2}$RuO$_4$ were grown by a floating-zone method \cite{Mao2000}.
We used the cleaved single crystals for the in-plane thermopower measurements, 
the photograph of which is shown in Fig. 2(a).
For the out-of-plane experiments,
the crystal was cut using an electrical discharge machine to obtain the elongated crystal  to the $c$-axis direction [Fig. 2(b)] \cite{Yamamoto2018}.
The thermopower was measured by a steady-state technique using two platinum resistance thermometers 
in a tube furnace \cite{Ikeda2016,Sakabayashi2021,Kurita2021}.
The thermoelectric voltage of the crystal was measured with a Keithley 2182A nanovoltmeter. 
The temperature gradient of about 0.5 K/mm was applied using a resistive heater. 
The thermoelectric voltage from the wire leads (platinum wires) was subtracted.

In order to verify how similar the band structure of Sr$_2$RuO$_4$ is to the band dispersion of the simple tight-binding model 
shown in Fig. 1(b), we also performed
first-principles calculations based on density functional theory (DFT)
using Quantum Espresso \cite{Giannozzi2009,Giannozzi2017,Giannozzi2020}.
We used the projector-augmented-wave pseudopotentials 
with the Perdew-Burke-Ernzerhof generalized-gradient-approximation (PBE-GGA) exchange-correlation functional.
The cut-off energies for plane waves and charge densities were set to 
70 and 560 Ry, respectively,
and the $k$-point mesh was set to $20\times 20 \times 20$ uniform grid
to ensure the convergence.
We used on-site Coulomb energy $U = 3.5$~eV and
 exchange parameter $J = 0.6$~eV for Ru ions \cite{Huang2020} and 
performed full relativistic calculations with spin-orbit coupling (${\rm DFT}+U+{\rm SOC}$).
The present calculations are not spin-polarized.

To examine the filtering effect,
we also calculated
the thermopower based on the linearized Boltzmann equations 
under constant relaxation time approximation using a Boltzwann module \cite{Giannozzi2014}.
From the obtained eigenvalues of the $n$-th band at $\bm k$ point $E_{n,\bm k}$,
the transport  function tensor $L_{ii}(\varepsilon)$ is calculated as
\begin{align}
L_{ii}(\varepsilon)
=
\sum_{n, \bm k}v_i^2\tau\delta(\varepsilon-E_{n,\bm k}),
\end{align}
where 
$v_i$ is the $i$-th component of the group velocity $\bm v = \frac{1}{\hbar}\nabla_{\bm k}E_{n,\bm k}$,
$\tau$ ($=10^{-14}$~s) is the relaxation time, and 
$\delta$ is the delta function  \cite{Madsen2006,Giannozzi2014}.
Here we consider the diagonal components $ii =xx$ and $zz$.
We then calculated the electrical conductivity tensor
$\sigma_{ii}(\mu) = e^2\int_{-\infty}^{\infty}d\varepsilon\left(-\frac{\partial f_0}{\partial \varepsilon}\right)L_{ii}$,
where $e$ is the elementary charge and $f_0$ is the Fermi-Dirac distribution function for 
the chemical potential $\mu$ and temperature $T$.
Similarly, the Peltier conductivity tensor is calculated as
$P_{ii}(\mu) = -\frac{e}{T}\int_{-\infty}^{\infty}d\varepsilon\left(-\frac{\partial f_0}{\partial \varepsilon}\right)(\varepsilon-\mu)L_{ii}$,
and the thermopower $S_{ii}$ is obtained as $S_{ii} = P_{ii}/\sigma_{ii}$.

\section{Results and Discussion}

\begin{figure}[t]
\begin{center}
\includegraphics[width=8.5cm]{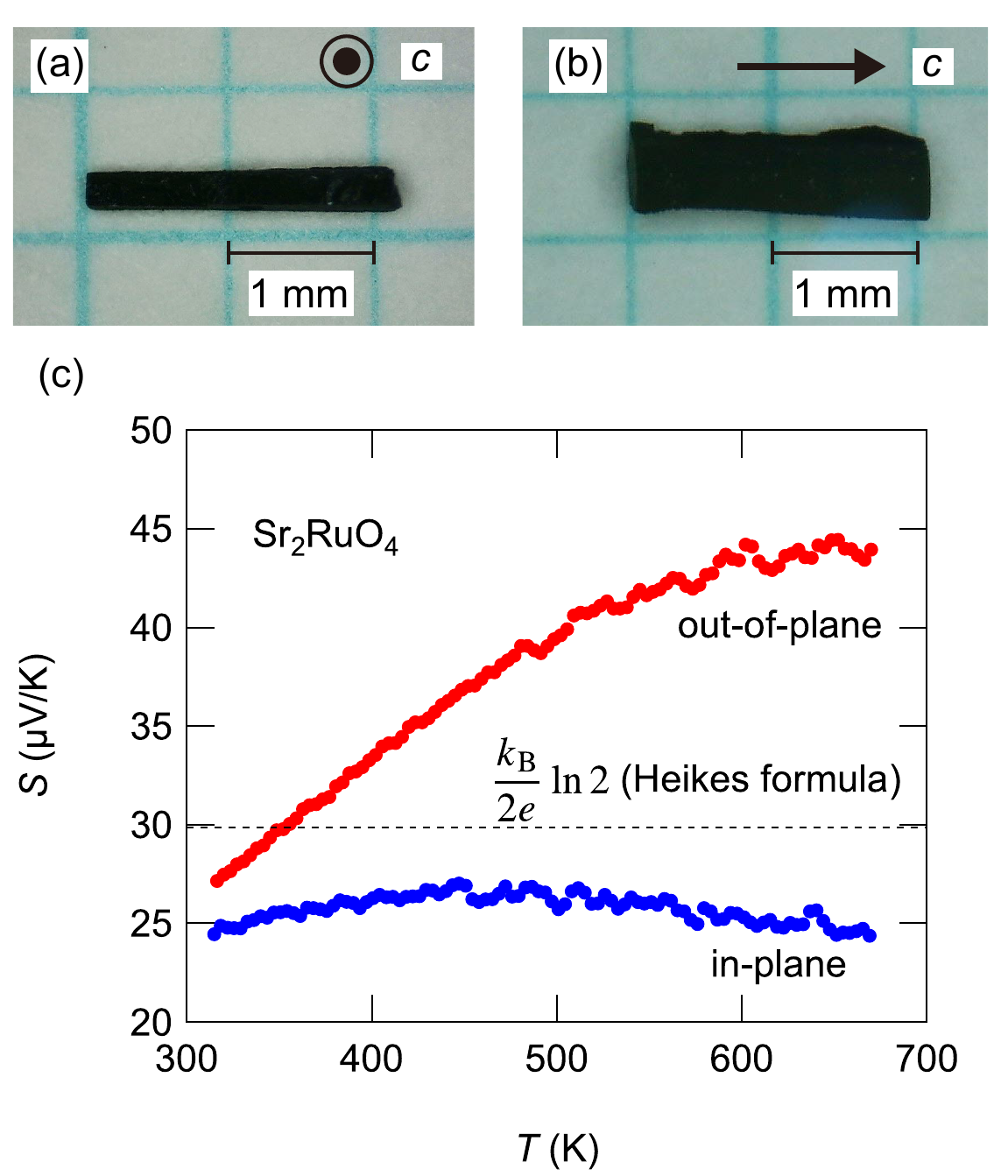}
\caption{
(a,b) Photograph of the single-crystalline Sr$_{2}$RuO$_4$ samples for the 
(a) in-plane and (b) out-of-plane thermopower measurements.
(c) Temperature dependence of 
the thermopower $S$ measured for the
in-plane (blue) and out-of-plane (red) directions.
The dashed line indicates a value expected from the Heikes formula at
high temperatures \cite{Mravlje2016}.
}
\end{center}
\end{figure}

\subsection{In-plane thermopower}

Figure 2(c) shows the temperature dependence of the 
in-plane (blue) and out-of-plane (red) thermopower.
We first discuss the in-plane thermopower behavior.
The room-temperature value of $S\sim25$~$\mu$V/K well
agrees with the earlier reports  \cite{Yoshino1996,Xu2008,Yamanaka2022},
in which the sample dependence was also observed in the magnitude of the thermopower.
Above room temperature, 
the in-plane thermopower exhibits weak temperature dependence,
which is similar to the thermopower measured in the polycrystalline samples \cite{Keawprak2008}.
This in-plane data
may be described by the Heikes formula for the mixed three valence states (atomic states with $N-1$, $N$, and $N+1$ electrons) \cite{Mravlje2016}
with the quenched orbital degeneracy due to the dominant Hund's coupling \cite{Georges2013,Karp2020}: 
\begin{align}
\label{Hek3v}
S = \frac{k_{\rm B}}{2e}\ln
\frac{g_{N-1}}{g_{N+1}},
\end{align}
where 
$k_{\rm B}$ is the Boltzmann constant
and
$g_i$ is the spin degeneracy of the atomic state with $i$ electrons.
For Sr$_{2}$RuO$_4$ ($N=4$), 
the Heikes thermopower is estimated as 
$S = k_{\rm B}/(2e)\ln(g_3/g_5) = k_{\rm B}/(2e)\ln 2\approx 30$~$\mu$V/K,
which is reasonably close to the experimental values.
Note that such concentration-independent formula may also be 
applicable to the high-temperature thermopower in the Mn oxides \cite{Kobayashi2004}.

\begin{figure}[t]
\begin{center}
\includegraphics[width=8cm]{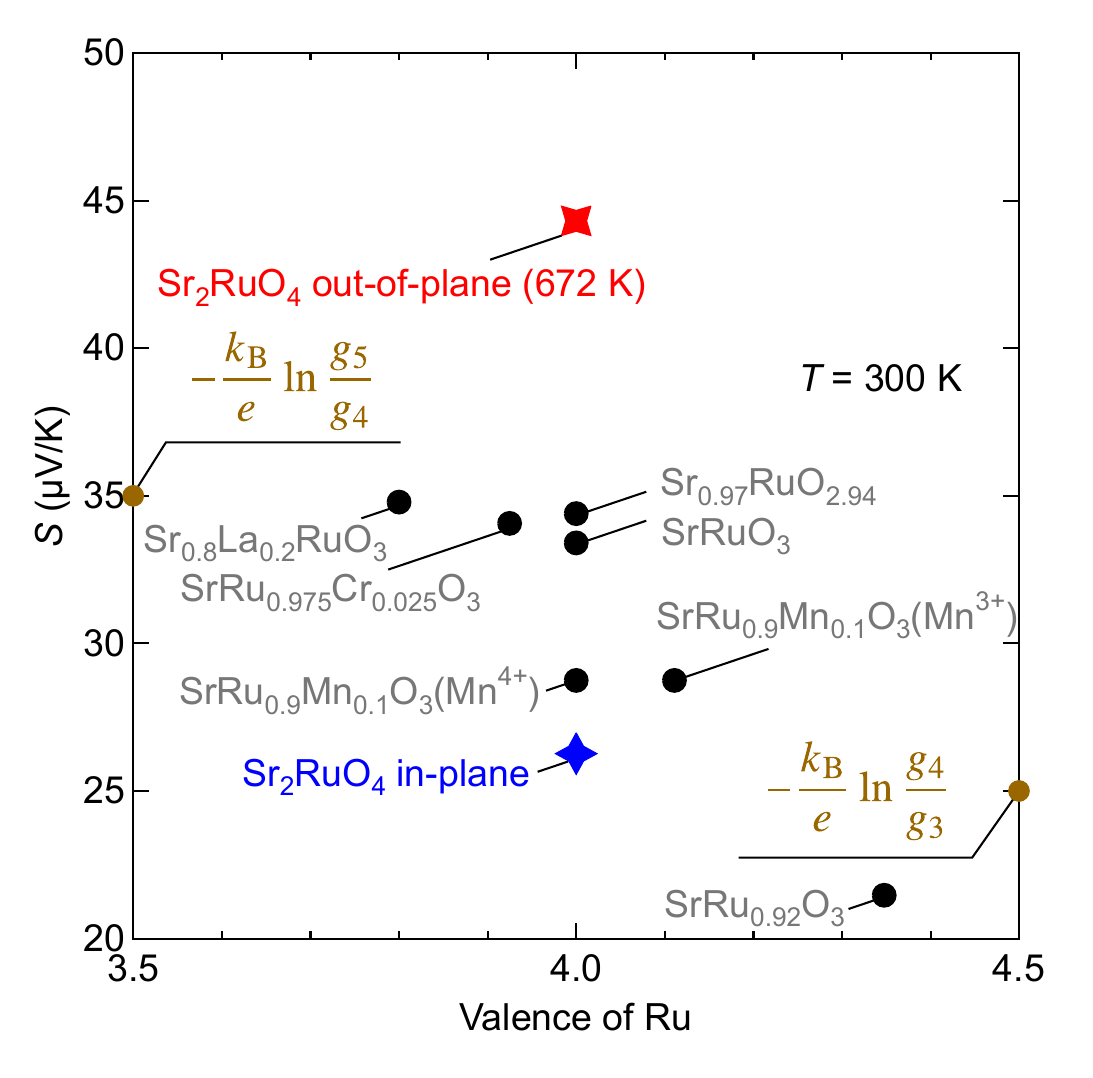}
\caption{
Thermopower in various itinerant ruthenium oxides
as a function of the formal valence of Ru ions.
The data were measured at room temperature 
except for the out-of-plane data of Sr$_{2}$RuO$_4$.
Data in the earlier reports are taken from Ref. \onlinecite{Klein2006}.
}
\end{center}
\end{figure}

Figure 3 summarizes the 
room-temperature thermopower data in various ruthenium oxides
as a function of the formal valence of Ru ions.
Note that we here focus on the in-plane data and the 
large out-of-plane thermopower at high temperatures 
will be discussed in the following section.
The thermopower in these ruthenium oxides is almost 
temperature-independent near room temperature \cite{Klein2006,Hebert2015},
similar to that of the present Sr$_2$RuO$_4$ for the in-plane direction.
In Fig. 3, 
the data seems to gather at the valence state of Ru$^{4+}$
with the thermopower value of $S=30$~$\mu$V/K as mentioned before.
In addition,
a trend of negative correlation between the thermopower value and the valence is obtained.
However, 
it is difficult to express the overall behavior by the Heikes formula:
Eq. (\ref{Hek3v}) is obtained by considering the mixed three valence states
and is valid only for $N=4$, whereas the extended Heikes formula
for mixed two valence states shows
divergent behavior of the thermopower close to the integer valence state \cite{Chaikin1976,Doumerc1994,Koshibae2000}.
For instance, 
in the valence range above 4 (mean number of electrons $n<4$),
the Heikes formula becomes 
\begin{align}
\label{Heik2v}
S = -\frac{k_{\rm B}}{e}\ln
\left(\frac{g_{\rm 4}}{g_{\rm 3}}\frac{4-n}{n-3}\right),
\end{align}
which diverges for the integer values of $n$.
Note that a mixed state with Ru$^{4+}$ ($N=4$) and Ru$^{5+}$ ($N=3$) ions is considered in Eq. (\ref{Heik2v}), 
in contrast to the situation for Eq. (\ref{Hek3v}) where three valence states are adopted.
In Fig. 3, we instead note the expected thermopower
of 
$(-k_{\rm B}/e)\ln(g_{4}/g_3)\approx25$~$\mu$V/K for 
the formal valence of 
Ru$^{4.5+}$ ($n=3.5$) and 
$(-k_{\rm B}/e)\ln(g_{5}/g_4)\approx35$~$\mu$V/K for 
Ru$^{3.5+}$ ($n=4.5$).
The data seems to locate between these values
and it is a future study to elucidate the exact formula to describe the thermopower
in the whole valence range.

\subsection{Out-of-plane thermopower}

\begin{figure}[t]
\begin{center}
\includegraphics[width=8.5cm]{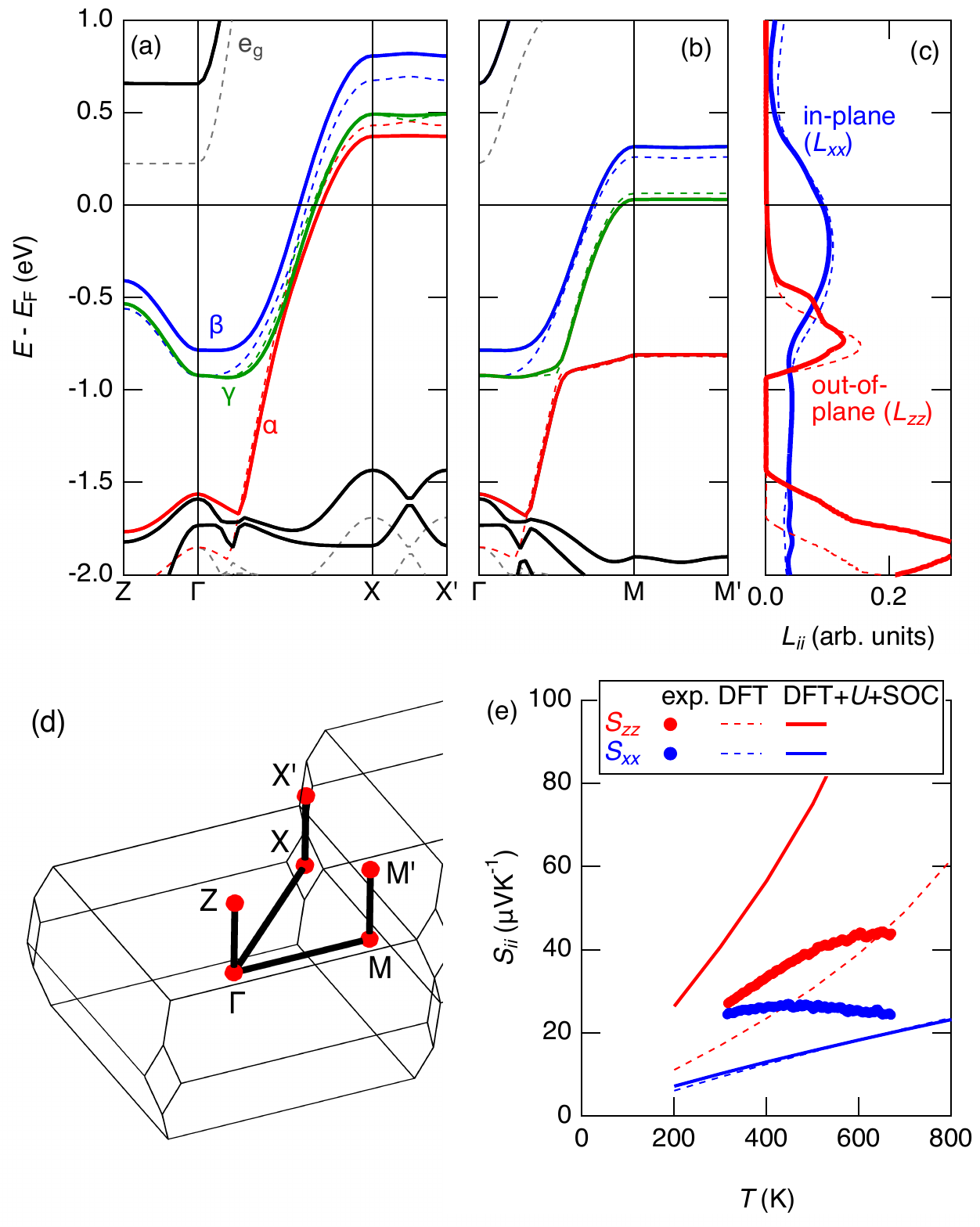}
\caption{
(a,b) Calculated band structure with ${\rm DFT}+U+{\rm SOC}$ scheme (solid curves)
along (a) the Z-$\Gamma$-X-X' and (b) the $\Gamma$-M-M' lines.
The dashed curves represent the results of scalar relativistic calculations and $U$ is not included.
The $\alpha$ (red), $\beta$ (blue), and $\gamma$ (green) bands cross the Fermi energy $E_{\rm F}$ indicated by 
the solid line.
(c) The in-plane and out-of-plane transport functions $L_{ii}$~($ii=xx,zz$). 
The solid and dashed curves represent the ${\rm DFT(GGA)}+U+{\rm SOC}$ and the DFT calculations, respectively.
(d) High-symmetry points and the $k$ path for the band structure in the panels (a) and (b).
The points X' and M' locate in the adjacent Brillouin zone.
(e) The calculated thermopower $S_{ii}$ for both directions as a function of temperature.
The solid symbols are experimental results.
The solid and dashed curves represent the ${\rm DFT}+U+{\rm SOC}$ and the DFT calculations, respectively.
}
\end{center}
\end{figure}

\begin{figure}[t]
\begin{center}
\includegraphics[width=8cm]{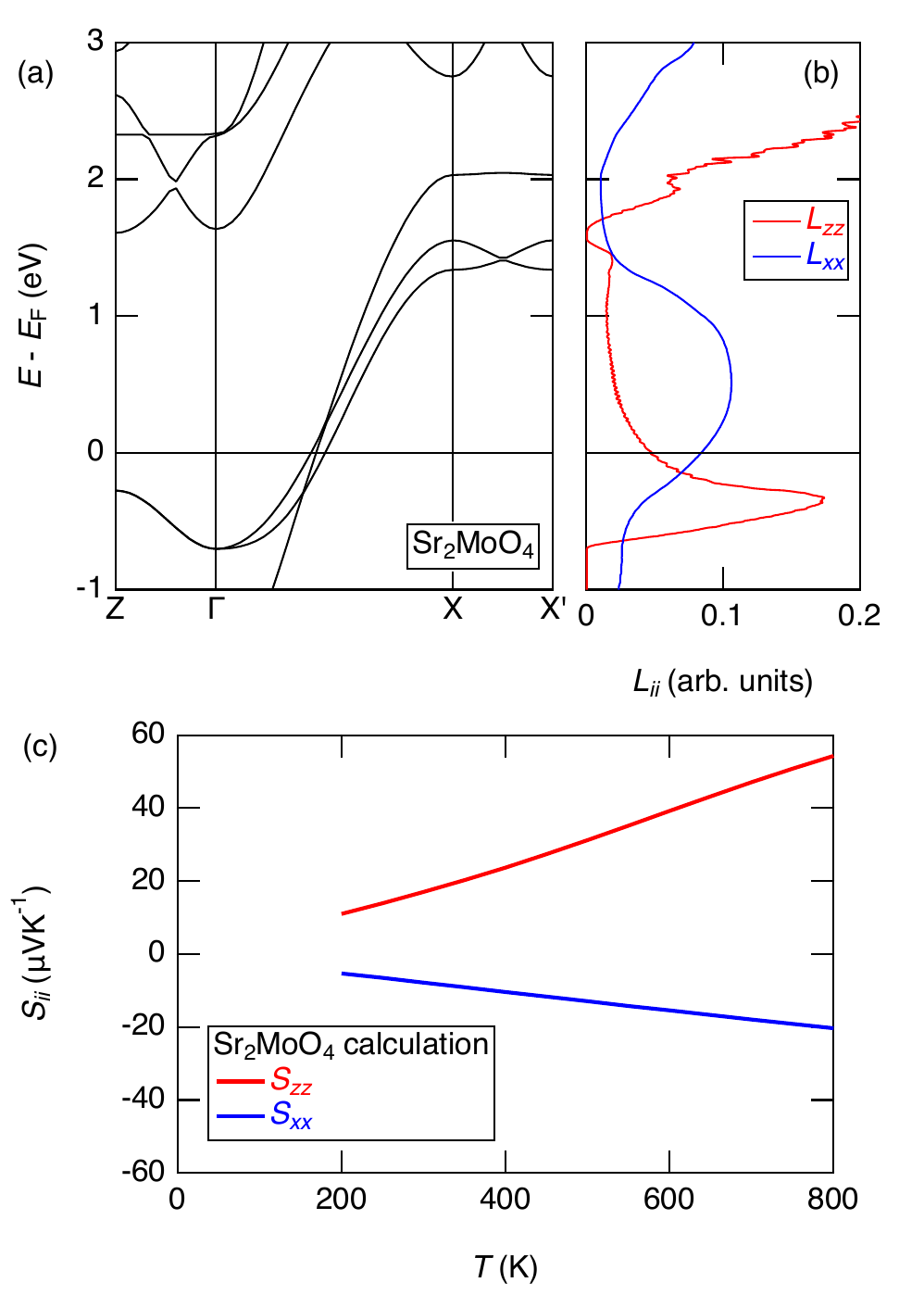}
\caption{
(a) Calculated band structure of Sr$_2$MoO$_4$ with GGA scheme
along the Z-$\Gamma$-X-X' line.
(b) The in-plane and out-of-plane transport functions $L_{ii}$~($ii=xx,zz$). 
(c) The calculated thermopower $S_{ii}$~($ii=xx,zz$) as a function of temperature.
}
\end{center}
\end{figure}

Then we focus on the out-of-plane thermopower,
which
increases with increasing temperature (Fig. 2)
and
well exceeds the above mentioned Heikes value of 30 $\mu$V/K (Fig. 3).
This behavior is highly distinct from the conventional trend:
as seen in the Heikes formula, the high-temperature thermopower will be expressed as 
the thermodynamic quantity with no directional anisotropy.
Indeed, for the layered cobalt oxides, 
the in-plane anisotropy in the thermopower gradually diminishes at higher temperatures \cite{Sakabayashi2021}.
Thus, it is clear that the Heikes picture is not applicable to 
the out-of-plane thermopower at high temperature range in Sr$_{2}$RuO$_4$. 
Rather, such a temperature dependence is well consistent with the 
theoretical estimation to consider the unique carrier filtering effect 
based on the bct lattice system \cite{Mravlje2016}.

To examine such filtering effect for the out-of-plane thermopower,
we calculate the band structure along the peculiar $k_z$ line
in the 
DFT(GGA) + $U$ + SOC scheme, 
since 
(i) the bandwidth may be corrected by the correlation term $U$ and
(ii) the inclusion of SOC may seriously affect the dispersion along the $k_z$ direction \cite{Haverkort2008},
which we focus here.
Figures 4(a) and 4(b) show the calculated electronic band structure near the Fermi energy $E_{\rm F}$, 
which well coincides with the results in earlier studies \cite{Oguchi1995,Singh1995,Hase1996,Haverkort2008}.
High-symmetry points and the $k$ path are shown in Fig. 4(d).
The solid curves represent the DFT+ $U$ + SOC  results, 
while the dashed curves are the conventional GGA results.
The $\beta$ and $\gamma$ bands at the high-symmetry points $\Gamma$ and Z split owing to 
the SOC \cite{Haverkort2008}, and the upper $e_g$ bands are shifted upward due to the on-site $U$.
Importantly, 
the $\beta$ and $\gamma$ bands along the 
Z-$\Gamma$-X-X' line are well reproduced with the tight-binding picture shown in Fig. 1(b):
the low-energy holes around $E-E_{\rm F}\approx-0.7$~eV possess the finite velocity along the 
Z-$\Gamma$ line.
In contrast, 
the $c$-axis velocity of the high-energy electrons around $E-E_{\rm F}\approx +0.5$~eV for the $\gamma$ band 
and $\approx +0.8$~eV for the $\beta$ band becomes almost zero due to the flat dispersion
along the X-X' line.
No $k_z$ dispersion is also observed at the edge of the Brillouin zone along the M-M' line [Fig. 4(b)].

Figure 4(c) shows the in-plane and out-of-plane transport function $L_{ii}$ $(ii=xx,zz)$ displayed in the same energy range for Figs. 4(a) and 4(b).
Corresponding to the band structure shown in Fig. 4(a), 
a significant peak structure of $L_{zz}$ is found near $E-E_{\rm F}\approx-0.7$~eV
in contrast to the negligibly small $L_{zz}$ for $E>E_{\rm F}$.
These prominent asymmetry in $L_{zz}$ owing to aforementioned large hole and small electron velocities
may  lead to
the enhanced positive thermopower at high temperatures.
Note that 
although the $L_{zz}$ peak is far from the Fermi energy ($E-E_{\rm F}\approx-0.7$~eV) 
compared to the thermal energy $k_{\rm B}T$ of the present measurement range,
this energy difference should become smaller by considering the scattering of correlated carriers accurately \cite{Mravlje2016}.
In contrast to highly asymmetric out-of-plane transport function, 
the in-plane one $L_{xx}$ in Fig. 4(c) seems symmetric around $E_{\rm F}$.
The thermopower data calculated with the different schemes are shown in Fig. 4(e).
Although the experimental data are not reproduced well due to the absence 
of mass renormalizations from the dynamical real-part of the self-energy \cite{Mravlje2016}, which is not included in the present DFT calculation, 
the enhanced out-of-plane thermopower $S_{zz}$ is obtained owing to the carrier filtering effect.
Interestingly, 
the out-of-plane thermopower calculated with DFT+$U$+SOC results
is much larger than that calculated with the DFT data. 
This difference originates from the 
upper shift of the  $\beta$ band due to the SOC \cite{Haverkort2008},
and the $L_{zz}$ peak near $E-E_{\rm F}\approx-0.7$~eV is slightly shifted to higher energy 
in the DFT+$U$+SOC results [Fig. 4(c)].
These results thus indicate that the SOC may affect the out-of-plane thermopower significantly.
Nevertheless, the calculated thermopower $S_{zz}$ from the DFT+$U$+SOC results is much larger than the 
experimental data of the out-of-plane thermopower.
This discrepancy may come from the absence of the self-energy analysis as described before.

We also mention other mechanisms to enhance the high-temperature thermopower.
Since this material shows an interesting bad-metallic transport at high temperatures \cite{Tyler1998},
the nature of the incoherent charge transport in such regime is of interest \cite{Hartnoll2014}.
In particular, the thermopower may be enhanced in the bad-metallic state near the Mott insular phase \cite{Zlatic2014},
whereas both in-plane and out-of-plane thermopower 
may be increased in such a case, which differs from the present results.

To explore the present carrier filtering effect in the bct lattice, 
we have calculated the electronic structure and the 
transport function for the related layered oxide Sr$_2$MoO$_4$ \cite{Ikeda2000}.
The calculation here is performed within the GGA scheme.
Figures 5(a) and 5(b) show the band structure, which well reproduce the earlier studies \cite{Hase2003,Karp2020},
and the transport function $L_{ii}$ for Sr$_2$MoO$_4$, respectively.
Similar to the results of Sr$_2$RuO$_4$, 
large difference in the out-of-plane velocities
between the Z-$\Gamma$ and the X-X' lines is observed for Sr$_2$MoO$_4$,
while the small $k_z$ dependence is also observed at around $E-E_{\rm F}\sim1.4$~eV.
Nevertheless, 
the out-of-plane transport function $L_{zz}$ is prominently asymmetric around $E_{\rm F}$ and exhibits a peak at around $E-E_{\rm F}\sim -0.3$~eV,
compared to relatively weak energy dependence of the in-plane transport function $L_{xx}$.
Figure 5(c) shows the calculated thermopower for Sr$_2$MoO$_4$.
Indeed, 
the out-of-plane thermopower $S_{zz}$ shows relatively large, positive values owing to the present hole filter effect.
In contrast, 
the in-plane thermopower $S_{xx}$ becomes negative 
because the electron number of the Mo ion is smaller than that in the Ru ion. 
Thus, Sr$_2$MoO$_4$ may also be interesting as a candidate of an oxide goniopolar material with axis-dependent polarity \cite{gonio}.
These results imply that the present carrier filtering 
may be applicable to 
wide range of the bct lattice such as K$_2$NiF$_4$-type or related structure  \cite{Ganguly1973,Matsuno2004,Pandey2013,Perry2006},
offering an interesting strategy toward the efficient thermoelectrics.

\textit{Note added.}
In the final stage of completion of this paper, 
we became aware of the preprint of Daou et al. \cite{arxiv},
which reports similar results on the anisotropic thermopower of Sr$_2$RuO$_4$ single crystals
at high temperatures.

\section{summary}

In summary,
we have measured the high-temperature thermopower of Sr$_{2}$RuO$_4$ single crystals 
for both in-plane and out-of-plane directions.
We find that the in-plane thermopower exhibits a weak temperature dependence,
which may be understood within the Heikes formula including the 
three different valence states.
Interestingly, 
the out-of-plane thermopower well exceeds the expected value of the Heikes formula
and increases with increasing temperature.
Such behavior may originate from the theoretically suggested carrier filtering effect,
which may be widely applied to various potential thermoelectrics with 
the body-centered tetragonal lattice system.

\section*{Acknowledgments}

We appreciate R. Nishinakayama, H. Shiina, and R. Taira for the assistance.
We thank the machine shop in Department of
Mechanical and Aerospace Engineering, Tokyo University of Science, for the use of electrical discharge machine.
This work was partly supported by JSPS KAKENHI Grant No. 17H06136 and No. 22H01166.


\end{document}